\documentclass[9pt,twocolumn,twoside]{osajnl}

\usepackage{amsmath,amsfonts,amssymb}
\usepackage{url}
\usepackage{bm, bbold}
\usepackage{color}
\usepackage{siunitx}
\usepackage{cuted}
\usepackage[normalem]{ulem}

\journal{ao} % Choose journal (ao, aop, josaa, josab, ol)

\setboolean{shortarticle}{false}
\ifthenelse{\boolean{shortarticle}}{\colorlet{color2}{color2b}}{\colorlet{color2}{color2}} % Automatically switches colors for short articles

\newcommand{\Ie}{I.\,e.}
\newcommand{\ie}{i.\,e.}

\newcommand{\eg}{e.\,g.}
\newcommand{\imag}{\mathrm{i}}

\newcommand{\dif}{\mathrm{d}}
\newcommand{\KK}{\mathcal{K}}

\newcommand{\cov}{\mathrm{cov}}
\newcommand{\var}{\mathrm{var}}
\def\Xint#1{\mathchoice
{\XXint\displaystyle\textstyle{#1}}%
{\XXint\textstyle\scriptstyle{#1}}%
{\XXint\scriptstyle\scriptscriptstyle{#1}}%
{\XXint\scriptscriptstyle\scriptscriptstyle{#1}}%
\!\int}
\def\XXint#1#2#3{{\setbox0=\hbox{$#1{#2#3}{\int}$}
\vcenter{\hbox{$#2#3$}}\kern-.5\wd0}}

\def\dashint{\Xint-}

\title{Determining the Refractive Index of Human Hemoglobin Solutions by Kramers-Kronig Relations with an Improved Absorption Model}

\author[1,*]{Jonas Gienger}
\author[1]{Hermann Gro\ss{}}
\author[1]{J\"{o}rg Neukammer}
\author[1]{Markus B\"{a}r}

\affil[1]{Physikalisch-Technische Bundesanstalt (PTB), Abbestra\ss{}e 2--12, 10587 Berlin, Germany}

\affil[*]{Corresponding author: jonas.gienger@ptb.de}

\dates{Compiled \today}

\ociscodes{
	  000.1430 Biology and medicine;
	  000.3860   Mathematical methods in physics;
	  120.4530 Optical constants;  
	  260.2030   Dispersion;
	  300.1030 Absorption
}

\doi{\url{http://dx.doi.org/10.1364/AO.55.008951}}

\begin{abstract}
The real part of the refractive index (RI) of aqueous solutions of human hemoglobin is computed
from their absorption spectra in the wavelength range \SI{250}{nm}--\SI{1100}{nm} using the Kramers-Kronig (KK) relations and the corresponding uncertainty analysis is provided.
The strong ultraviolet (UV) and infrared absorbance of the water 
outside this spectral range were taken into account in a previous study employing KK relations. We improve these results by including the concentration dependence
of the water absorbance as well as by modeling the deep UV absorbance of hemoglobin’s peptide backbone.
The two free parameters of the model for the deep UV absorbance are fixed by a global fit.
\\{\footnotesize \normalfont © 2016 Optical Society of America. One print or electronic copy may be made for personal use only. Systematic reproduction and distribution, duplication of any material in this paper for a fee or for commercial purposes, or modifications of the content of this paper are prohibited.}
\end{abstract}

\setboolean{displaycopyright}{false}

\begin{document}

\maketitle
\thispagestyle{fancy}

\ifthenelse{\boolean{shortarticle}}{\ifthenelse{\boolean{singlecolumn}}{\abscontentformatted}{\abscontent}}{}

\section{Introduction}\label{sec:intro}
The optical properties of biological cells and tissues have been subject to research for many decades. 
Refractometry in cells  is used, \eg, for protein detection \cite{Barer1954refracometry}.
Recently, the refractive index (RI)  or RI distribution  of cells has been measured
by phase microscopy \cite{Park2009spectroscopic}, holographic techniques \cite{Sung2014holographic}, absorption cytometry \cite{Schonbrun2014absorptioncytometry}
or optical tomography \cite{Kim2014profiling}.
From such measurements, the dry mass or concentrations of proteins in the cell can be derived, provided the relation to the optical properties is known.

Analysis of blood samples includes the determination of the quantities of the so called complete blood count (CBC), 
one of the most frequently performed measurements in laboratory medicine. Besides the concentrations of red blood cells (RBCs, erythrocytes), 
white blood cells and platelets, an important indicator for diseases, \eg, anemia, is the mean cellular volume (MCV) of red blood cells. 
Besides the calculation of the MCV as ratio of the hematocrit value and the RBC concentration, flow-cytometric detection of light scattering by
RBCs is being used for more than three decades to determine the volume and hemoglobin content of individual cells \cite{tycko1985flow, Mohandas1986independent} at a throughput in the range of 1000 events per second.
Of course, a detailed knowledge of the cells' refractive index is required for this.  
Since RBCs are mainly composed of the protein complex hemoglobin (Hb), which is dissolved in water
at an average intracellular concentration of typically \SI{340}{g.L^{-1}} for healthy persons,
the knowledge of the Hb RI is of crucial importance to reliably calculate RBC volume and Hb content.

Within the blood of a single person, the intracellular Hb concentration fluctuates among the individual RBCs, with a
coefficient of variation around 6--8\% \cite{tycko1985flow, Kim2014profiling}.
Because of the significant influence of this variation,  
it does not suffice to know the RI of hemoglobin solutions at the mean concentration, but the
dependence of the RI on the concentration has to be specified.

The absorption spectra of Hb solutions were measured with high accuracy over a wide range of wavelengths
and are known for several decades. 
In contrast, measurements of their refractive index, especially at physiologically relevant high concentrations, are challenging and have 
only been presented as late as 2005  for  a wide spectral range \cite{Friebel2005concentratedHb}. However, these data have much larger
measurement uncertainties than the corresponding absorption spectra, such that the real part of the complex refractive index is
 less accurately known than its imaginary part, the absorbance.

As a remedy several authors \cite{faber2004oxygen, Sydoruk2012refractiveindex}  suggested to employ Kramers-Kronig (KK) relations in order to obtain the
real part of the RI directly from accurate measurements of  absorption spectra, \ie,
the imaginary part of the RI as a function of wavelength.
However, this is a non-trivial task due to the finite wavelength range of the measured spectra and the global, long-ranged character
of the KK relations which yield the real part of the RI as an integral transformation of the imaginary part.
In this paper, we supplement and extend the literature spectra \cite{Friebel2005concentratedHb}
with an absorption model for proteins towards the UV.
As a result we obtain RIs that are in better agreement with results obtained by reflectance measurements for oxyhemoglobin solutions \cite{Friebel2006modelfunction}.
Furthermore we obtain results for the RI of deoxyhemoglobin
and give the \emph{refractive increment}, \ie, the slope of the RI with respect to concentration  for the first time.
We also extend the previous treatment by considering the measurement uncertainties of the obtained RIs.
These uncertainties result from the propagation of the uncertainties of the measured absorption spectra  \cite{Friebel2005concentratedHb} and the
refractive index data obtained by reflectance measurements \cite{Friebel2006modelfunction}, to which we fit the two free parameters of the
deep UV absorption model.
The mathematical model for the absorption spectra and related method are presented in sections~\ref{sec:methods} and \ref{sec:results}.
Our approach can be applied to any variant of hemoglobin.
Section~\ref{sec:methods}\,\ref{sec:spectra} describes the imaginary part of the RI of a Hb solution in dependence on concentration.
KK relations are  used  in section~\ref{sec:methods}\,\ref{sec:realRI} to obtain an expression for the
real part of the solution's RI. A  model for the deep UV absorbance of hemoglobin is proposed in section~\ref{sec:methods}\,\ref{sec:realRI} as well.
Section~\ref{sec:methods}\,\ref{sec:numeval} describes the fitting of the two free model parameters to the literature data for the real part of the RI.
In section~\ref{sec:results}, we apply the analysis to experimental data for oxygenated and deoxygenated hemoglobin, respectively.
The significance of the results is discussed and a comparison to previous KK-analyses is made. In section~\ref{sec:summary_outlook}, we summarize our findings.

\subsection*{Terminology}
Oxygenated hemoglobin is often abbreviated as ``$\mathrm{HbO_2}$'' while deoxygenated
hemoglobin is usually abbreviated as ``Hb''. 
In order to simplify the notation we will use the term ``Hb'' for hemoglobin in general, specifying the hemoglobin variant only if relevant.

The RI of Hb solutions complex-valued. For the sake of brevity, we 
refer to the real part of the RI by ``real RI'' and to the imaginary part of the RI by ``imaginary RI''.
This shall not imply that the RI is purely real or imaginary.

\section{Materials and Methods}\label{sec:methods}
\subsection{Absorption Spectra: Imaginary Part of Refractive Index}\label{sec:spectra}
The    Hb absorption spectra in the wavelength range $[250,1100]\,\si{nm}$
reported in \cite{Friebel2005concentratedHb} were used in our calculations.
These data were measured for
Hb solutions produced from human erythrocytes by repeated freezing and thawing followed by centrifugation of the membrane components.
These homogeneous solutions did thus contain all the constituents of 
human erythrocytes, except for the membranes (thickness less than a few \SI{10}{nm}, volume fraction less than a few\,\% \cite{deuling1976RBC,tycko1985flow,heinrich2001elastic}).
We further use experimental data for pure Hb solutions to supplement the absorbance data in the wavelength range \SI{228}{nm}--\SI{250}{nm}  \cite{Sugita1971circular}.

Other researchers \cite{faber2004oxygen, Sydoruk2012refractiveindex} have previously relied on the 
data compiled from various sources by Prahl \cite{prahlHbspectra}.
However, since not all of the compiled sources used human hemoglobin, these data are  not used  here.

\subsubsection{Literature Data}\label{subsec:literatureabsorption}
\begin{figure}[t]
 \centering
 \setlength{\fboxsep}{0pt}
 \includegraphics[width=\linewidth]{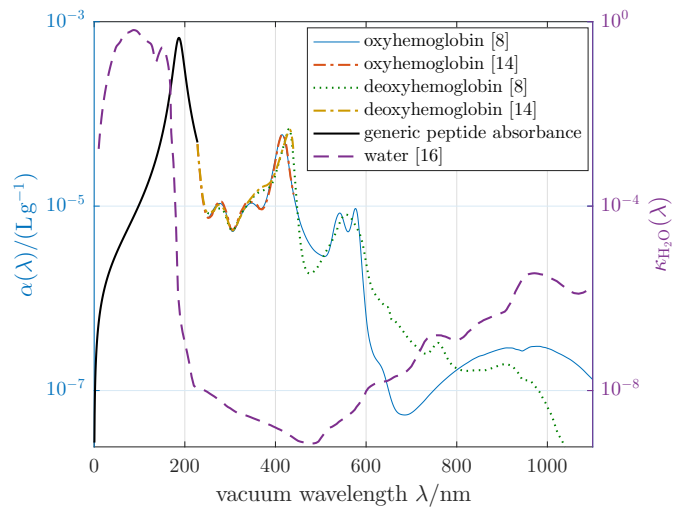}
 \caption{
  Imaginary refractive increment 
  $\alpha(\lambda)$ of hemoglobin in aqueous solutions (left axis, solid and dash-dotted lines).
  Experimental data are taken from \cite{Friebel2005concentratedHb} (corrected for ${\rm H_2O}$ absorbance) and \cite{Sugita1971circular}.
  A Lorentzian peak  models the deep UV absorbance of the peptide backbone (left axis, solid black line).
  Imaginary RI $\kappa_{\rm H_2O}(\lambda)$ of water (right axis, dashed line) \cite{segelstein1981complex}. Note that the quantities $\alpha, \kappa_{\rm H_2O}$ have different units and scaling of $y$-axes.
  To compare the numerical values of $\alpha$ and $\kappa_{\rm H_2O}$, one needs to multiply $\alpha$ by the respective concentration of the RBC, \eg, $c_{\rm Hb} = \SI{340}{g.L^{-1}}$, cf. \eqref{eq:def_alpha}.
  Spline interpolation was applied to obtain a step width of \SI{1}{nm}.}
 \label{fig:spectra}
\end{figure}
The literature absorption spectra we use in our calculations are expressed in terms of the inverse absorption length $\mu_a(\lambda)$ at a given mass  concentration $c$  or
the molar extinction coefficient $\varepsilon_M(\lambda)${, where $\lambda$ is the vacuum wavelength of the light}. The latter implies that the  Lambert-Beer law holds, which
is  the case for Hb at physiological ($c\approx \SI{300}{g.L^{-1}}%, c_M\approx \SI{4.7}{\milli\mole.L^{-1} }
$) or lower concentrations \cite{Friebel2005concentratedHb}. 
Instead of $\mu_a(\lambda)$ or $\varepsilon_M(\lambda)$, we express the absorption spectra in terms of the imaginary part  of the complex RI
\begin{equation}
 \mathfrak{n}(\lambda) = n(\lambda) + \imag\,\kappa(\lambda)
\end{equation}
of the solution.
Here $n, \kappa$ are positive real functions.
The conversion rule
is
\begin{equation}
\kappa(\lambda) =  \frac{\mu_a(\lambda)\,\lambda}{4\pi} = \frac{\ln10\,\varepsilon_M(\lambda)\,c\,\lambda}{4\pi\, M}. \label{eq:conversion_kappa_mu}
\end{equation}
The data on the molar extinction coefficient $\varepsilon_M(\lambda)$ from \cite{Sugita1971circular} are converted to $\kappa(\lambda)$, using the molar mass of the Hb tetramer $M = \SI{64458}{g.mol^{-1}}$ \cite{braunitzer1964molecular}.

\subsubsection{Total Absorbance of Hb Solutions and Erythrocytes}
In the spectral range of $\lambda\in[250,1100]\,\si{nm}$ the strongest absorption of light by  RBCs and hence of blood is caused by Hb.
Water has a fairly low absorption coefficient in this region (cf. Fig.~\ref{fig:spectra}) and other RBC components contained in the cytosol (\eg, other proteins, sugars, ions)
exhibit rather low concentrations.
We thus consider a two-component system, \ie,
\begin{enumerate}
 \item the solvent, or simply ``Water'' ($\rm H_2O$), occupying a volume $V_{\rm H_2O}$ and
 \item the absorbing solute, or simply ``Hemoglobin'' (Hb), denoting everything else contained in the erythrocyte cytosol 
  and occupying a volume $V_{\rm Hb}$.
 The solute is treated as pure Hb for its other physical properties, such as molar mass and density.
\end{enumerate}

Let the volume fraction occupied by Hb molecules be 
$
  { \phi = V_{\rm Hb}/({V_{\rm H_2O}+V_{\rm Hb}}})
$
and the volume fraction occupied by water molecules $1-\phi = { V_{\rm H_2O}/({V_{\rm H_2O}+V_{\rm Hb}}})$.
Since the Lambert-Beer law holds, we make the following ansatz for the total imaginary RI
(absorbance) of the solution
\begin{equation}
 \kappa(\lambda) = \phi\,\kappa_{\rm{Hb}}(\lambda) + (1-\phi)\,\kappa_{\rm{H_2O}}(\lambda), \label{eq:ansatz_absorption}
\end{equation}
where $\kappa_{\rm H_2O}(\lambda)$  is the absorbance of pure water and $\kappa_{\rm Hb}(\lambda)$ is the absorbance of ``pure hemoglobin in aqueous solution''.
It should be noted that $\kappa_{\rm Hb}$ is not equal to the imaginary RI of a crystal of pure hemoglobin as is revealed by comparison with
data for the complex dielectric function of thin Hb films \cite{arwin1986albumin}.
\eqref{eq:ansatz_absorption} is similar to the ansatz in \cite{Sydoruk2012refractiveindex}, where, however, the authors did
not include the prefactor $1-\phi$ for the water absorbance, which results in a relevant difference as shown later. 
This \emph{excluded volume} effect,
\ie, the fact that there is less water in a solution of higher Hb concentration is essential and
must be taken into account since the volume fractions of Hb can be as high as 26\% or more.

The Hb volume fraction is related  to the concentration by
\begin{equation}
 \phi = \frac{c_\text{Hb}}{\rho_{\rm Hb}}, 
\end{equation}
where $\rho_{\rm Hb}$ is the mass density of a hypothetical solution containing 100\%\,Hb and 0\%\,$\mathrm{H_2O}$. Both, $\kappa_{\rm{Hb}}(\lambda)$ and $\rho_{\rm{Hb}}$
are coefficients in a linear interpolation of spectra and density that holds at least up to physiologically high concentrations.  
A  value of $\rho_{\rm Hb} = 1330\,\mathrm{g\,L^{-1}}$ for proteins is given in \cite{Barer1954refracometry}, independent
of the type of protein.
A normal physiological intra-erythrocyte Hb concentration of $340\,\mathrm{g\,L^{-1}}$ thus corresponds to a volume fraction of
$\phi\approx0.26$ or $26\%$.

Then the term for the absorbance contribution of Hb, $\phi\,\kappa_{\rm Hb}(\lambda)$, becomes
\begin{equation}
 \phi\,\kappa_{\rm Hb}(\lambda) = c_\text{Hb}\frac{\kappa_{\rm Hb}(\lambda)}{\rho_{\rm Hb}} \;\stackrel{\text{def. } \alpha(\lambda)}{=}\; c_\text{Hb}\,\alpha(\lambda), \label{eq:def_alpha}
\end{equation}
where $\alpha(\lambda)$ is Hb's concentration-specific increment of the imaginary RI, or the \emph{imaginary refractive increment}.  Since 
the Hb concentration $c_{\rm Hb}$ is  measured in $\si{g.L^{-1}}$, the unit for $\alpha$ is $\si{L.g^{-1}}$.
\eqref{eq:ansatz_absorption}   becomes
\begin{equation}
 \kappa(\lambda) = c_{\rm Hb}\,\alpha(\lambda) + \left(1-\frac{c_{\rm Hb}}{\rho_{\rm Hb}}\right)\,\kappa_{\rm{H_2O}}(\lambda). \label{eq:ansatz_absorption_2}
\end{equation}

\eqref{eq:ansatz_absorption_2} together with  \eqref{eq:conversion_kappa_mu} and \eqref{eq:ansatz_absorption} allows to compute the imaginary refractive increment $\alpha(\lambda)$ from 
{ a measurement of} the inverse absorption length $\mu_a^*(\lambda)$ of a solution at known concentration $c_{\rm Hb}^*$ as
\begin{align}
 \begin{split}
 \alpha(\lambda) &= \frac{1}{c_{\rm Hb}^*}\left[  \kappa^*(\lambda) + (\phi^*-1)\,\kappa_{\rm H_2O}(\lambda)  \right] \\
  &= \frac{1}{c_{\rm Hb}^*}\,\frac{\lambda}{4\pi}\left[ \mu_a^*(\lambda) + \left(\frac{c_{\rm Hb}^*}{\rho_{\rm Hb}}-1\right)\,\mu_{a,{\rm H_2O}}(\lambda)  \right],
 \end{split}
\end{align}
where the asterisk $^*$ denotes experimental data.
The inverse absorption length of water $\mu_{a,{\rm H_2O}}(\lambda)$ is known to high accuracy over a large spectral range $\lambda \in[ \SI{10}{nm}, \SI{10}{m}]$ \cite{segelstein1981complex}. Hence this formula allows to 
correct for the water absorption, which is  important in the infrared (IR), where Hb absorbs only weakly.

\subsection{Real Part of  Refractive Index by Kramers-Kronig Relations}\label{sec:realRI}

\subsubsection{Literature Data}
The most complete experimental measurement of the wavelength-dependent real RI of Hb solutions to date was presented in \cite{Friebel2005concentratedHb, Friebel2006modelfunction}.
Here, $n(\lambda)$ was determined via measurements of the spectral reflectance $R(\lambda)$ at an interface between air and Hb solution at normal incidence.
The reflectance at normal incidence is connected to the complex refractive index by the Fresnel equation
\begin{equation}
 R(\lambda) = \left|\frac{\mathfrak{n}(\lambda)-1}{\mathfrak{n}(\lambda)+1}\right|^2 = \frac{(n(\lambda)-1)^2 + \kappa(\lambda)^2}{(n(\lambda)+1)^2 + \kappa(\lambda)^2},
 \label{eq:Fresnel}
\end{equation}
which is easily solved for $n(\lambda)$ when $\kappa(\lambda)$ is known or when  it can be neglected  because $\kappa(\lambda)\ll n(\lambda)-1$.

In \cite{ Friebel2006modelfunction} measurements at different concentrations were  analyzed. A linear-affine dependence of the real RI on the concentration was found
and the result was expressed as
\begin{equation}
 n(\lambda) = n_{\rm H_2O}(\lambda) \, [1+c_{\rm Hb}\, \beta(\lambda)]. \label{eq:FriebelModel}
\end{equation}
Here $\beta(\lambda)$ is the concentration-specific  increment of the real RI relative to the water RI.

\subsubsection{Kramers-Kronig Relations}
The complex refractive index is a linear, causal response-function to an incident wave.
Hence its real and imaginary part are  related by KK relations. 
Expressed in terms of wavelengths, the KK relations  for the complex RI read
\begin{align}
  n(\lambda) - 1 = \KK[\kappa](\lambda)&:= -\frac{2}{\pi} \dashint_0^\infty \frac{\lambda}{\Lambda} \frac{\lambda}{\Lambda^2-\lambda^2}  \,\kappa(\Lambda)  \, \dif \Lambda,  
  \label{eq:KK_nRe}\\  
  \kappa(\lambda) = \KK^{-1}[n](\lambda)&\;= +\frac{2}{\pi} \dashint_0^\infty  \frac{\lambda }{\Lambda^2-\lambda^2}  \,n(\Lambda)  \, \dif \Lambda,   \label{eq:KK_nIm}
\end{align}
where \eqref{eq:KK_nRe} also defines the integral transform $\KK$ in general. The symbol $:=$ denotes a definition and  the symbol $\dashint$ denotes the Cauchy principal value integral.

Applying \eqref{eq:KK_nRe} formally to the ansatz for the absorption of the Hb solution \eqref{eq:ansatz_absorption_2}, we obtain
\begin{equation}
 n(\lambda) -1 = c_{\rm Hb}\,G(\lambda) + \left(1-\frac{c_{\rm Hb}}{\rho_{\rm Hb}}\right)\,\left(n_{\rm{H_2O}}(\lambda)-1\right), \label{eq:n_Re_from_ansatz}
\end{equation}
where $G(\lambda):= \KK[\alpha](\lambda)$ is the transformed spectrum (cf. \eqref{eq:KK_nRe}).
This formal transformation of the absorption  results in an equation for the real RI of the Hb solution
\begin{align}
\begin{split}
 n(\lambda) &= n_{\rm H_2O}(\lambda) +
 c_{\rm{Hb}}\,\left[G(\lambda) - \frac{n_{\rm H_2O}(\lambda) - 1}{\rho_{\rm{Hb}}} \right]\\
 & {\hspace{-8pt}\stackrel{\text{def. }B(\lambda)}{=}}
 n_{\rm H_2O}(\lambda) + c_{\rm{Hb}}\,B(\lambda). \label{eq:realRI_linear}
 \end{split}
\end{align}

The linear-affine %(or rather, affine) 
dependence of $n(\lambda)$ on $c_{\rm Hb}$ in \eqref{eq:realRI_linear} is  in agreement with experimental findings \cite{Friebel2006modelfunction} (cf. \eqref{eq:FriebelModel}).
In \eqref{eq:ansatz_absorption_2}, we have formally split off the water absorption, such that $n_{\rm H_2O}(\lambda)$ 
contributes to the background in the dispersion  relations of the Hb solutions (cf. \eqref{eq:realRI_linear}).
Since the real RI of water -- unlike the real RI of Hb -- is known to high accuracy, this
provides  valuable additional information  compared to the application of the KK-transform only to the
measured absorption spectrum of a Hb solution in the visible and near UV/IR range that was presented in \cite{faber2004oxygen}.
This idea was already presented in \cite{Sydoruk2012refractiveindex}.
However, due to our different ansatz for the absorption, where we take into account the \emph{excluded water volume}, we obtain a different result for $B(\lambda)$
with an additional term $(n_{\rm H_2 O}(\lambda) -1)/\rho_{\rm Hb}$ for the concentration dependence.
We discuss the differences between the  results obtained in  \cite{faber2004oxygen}, \cite{Sydoruk2012refractiveindex} and 
 the result with our improved model  in section \ref{sec:results}\,\ref{subsec:comparison_previous_KK}.

For numerical values of $n_{\rm H_2O}(\lambda)$, we use a  four-term Sellmeier formula that is accurate  to  at least five decimal places
\cite{Daimon2007refractivewater}.

\subsubsection{Additional Spectral Information Outside the Measured Range}
The KK relations provide a formal tool to derive results like \eqref{eq:realRI_linear}.
However, their application has  a  well known  problem for the numerical evaluation:
They are global integral transforms that require the knowledge of real or imaginary RI at all wavelengths $\lambda \in [0,\infty[$, 
which is practically impossible.
Although the integral kernel in \eqref{eq:KK_nRe} is decaying  with increasing distance from the pole at $\Lambda=\lambda$,
it is long-ranged. Hence, one cannot simply cut off the integration domain, \ie, use a finite dataset.

Water is transparent  to visible light and its main regions of absorption lie  in the UV and the IR (Fig.~\ref{fig:spectra}).
Due to these absorption bands water has a RI significantly different from 1, even in the transparent regions and exhibits normal dispersion,
\ie, $n(\lambda)$ decreases with $\lambda$. 
This implies that neglecting such contributions in the KK relations   may   lead to inaccurate results.
We will now  describe the absorption features of Hb below \SI{250}{nm} by a mathematical model  and include them in our expression for the real RI.

In addition to the known absorption spectrum (Fig.~\ref{fig:spectra}) with strong absorption { in the vicinity of
the Soret band at  } \SI{420}{nm},  Hb has an  even stronger absorption peak in the deep UV.
This feature stems from the peptide bonds forming the backbone of any polypeptide or protein, including the protein complex hemoglobin
and is characteristic for  polypeptides and proteins.
The corresponding extinction coefficient curves
$\varepsilon(\lambda)$ are similar among a variety of proteins
and the absorbance maximum is typically located at $\lambda = \SI{187}{nm}$  \cite{Goldfarb1951UVproteins,Woods1970absorptionproteins}.

This  peptide-peak must be accounted for  to perform a proper KK analysis, but is, unfortunately, not resolved in the existing  experimental  Hb spectra.
However, data were reported for human and bovine albumin  \cite{Woods1970absorptionproteins} 
-- a protein found in blood serum. Albumin  is  similar to hemoglobin in its mass and optical properties
at wavelengths away from the characteristic  Hb absorption band at $\SI{420}{nm}$.
The absorption maximum for human albumin is reported as $\varepsilon(\SI{187}{nm}) = \SI{86.0}{L.g^{-1}.cm^{-1}}$,
corresponding to a value of $\alpha(\SI{187}{nm}) = \SI{2.95e-4}{L.g^{-1}}$, which is more than four times as high as
the peak around \SI{420}{nm} (Fig.~\ref{fig:spectra}).

We model this generic protein absorption using an anti-symmetrized Lorentzian curve 
\begin{equation}
 \alpha_L(\lambda) = a_L\frac1\pi \frac{\Gamma}{(\lambda-L)^2+\Gamma^2} - a_L\frac1\pi \frac{\Gamma}{(\lambda+L)^2+\Gamma^2}, \label{eq:antisymm_Lorentzian}
\end{equation}
where $\Gamma = \SI{11.6}{nm}$ is the { half width at half maximum } of the curve and $L=\SI{187}{nm}$ is the position of its maximum. 
This model curve fulfills $\alpha_L(-\lambda) = -\alpha_L(\lambda)$. This is important,
 as { the symmetries} $\kappa(-\lambda) = -\kappa(\lambda)$ and $n(-\lambda) = n(\lambda)$ are implied when using the KK relations in the form \eqref{eq:KK_nRe}, \eqref{eq:KK_nIm} or
the analogous expressions for the frequency,  where they are written as an semi-infinite integral, denoted by the symbol $\dashint_0^{\infty}$.

As mentioned before,  the deep UV spectrum and hence the  exact shape of the peptide absorption line is not available in the literature.
For proteins, Woods and O'Bar  report that ``the increase in absorbance at $\SI{187}{nm}$ is threefold over that at \SI{205}{nm} and fourfold
over that at \SI{210}{nm}''\cite{Woods1970absorptionproteins}. This description fits well to  the half width of the curve of
$
 \Gamma = \SI{11.6}{nm}.
$
Going to even lower wavelengths $\lambda\ll\SI{187}{nm}$  there will be more absorption features, since  the inner electron shells of the atoms will be excited. 
We thus expect a variety of overlapping absorption  lines  at these  short wavelengths.
On the other hand, this spectral region is fairly far away from the region of interest and
the KK relations contain a damping factor of $1/(\lambda^2-\Lambda^2)$. Hence, the exact line-shapes are  practically irrelevant. 
For our model, it suffices to add  to the spectrum 
a delta-peak of unknown amplitude  located at zero wavelength,
which accounts for the influence of extreme UV absorption by a constant offset in the real RI:
$\alpha_\delta(\lambda) = \lim_{\lambda_\delta \to 0^+}\frac{\pi}{2}\,a_\delta\,\lambda_\delta\,\delta(\lambda - \lambda_\delta)$.

In this respect, our approach is not different from \cite{faber2004oxygen, Sydoruk2012refractiveindex}: 
We cannot predict the absolute value of the 
RI of Hb, but we need to  determine  a constant by comparing to experimental data.
However, we apply non-local fitting to optimize this constant, instead of using just one single data point.
Thus, our method is more robust to uncertainties in both the absorption spectra and the measured real RIs.

\subsection{Fitting to Literature Values}\label{sec:numeval}
With this model for the UV absorption, we have an absorption spectrum
\begin{equation}
 \alpha(\lambda) = \alpha_\text{lit}(\lambda) + \alpha_L(\lambda) +  \alpha_\delta(\lambda),
\end{equation}
where $\alpha_\text{lit}(\lambda)$ represents the experimental literature data.
For the integral transform $\KK$ (cf. \eqref{eq:KK_nRe}) the
contribution from the  $\delta$-peak
$\alpha_\delta(\lambda)$
 is $  G_\delta(\lambda) = a_\delta$ and hence constant
and the contribution $G_L(\lambda) = a_L\,\widetilde{G}_L(\lambda)$ from the Lorentzian, integrated only over the deep UV part of the spectrum, can be obtained analytically (see appendix~\ref{app:trafo_Lorentzian}).
Here $\widetilde{G}_L(\lambda)$ is the contribution for a Lorentzian of unit amplitude.
Thus, we are left with
\begin{equation}
 G(\lambda) = G_\text{lit}(\lambda) + a_L\,\widetilde{G}_L(\lambda) +  a_\delta,
\end{equation}
where only the first term $ G_\text{lit}(\lambda):=\KK[\alpha_\text{lit}](\lambda)$  needs to be evaluated numerically.

Numerical evaluation is straightforward. 
We use an integration scheme,
which evaluates the KK relations as a Riemann sum with Taylor expansion at the singularities of the integrand
as described, \eg, in \cite{Emeis1967numericalKK} and in appendix~\ref{app:num_int}.
The experimental literature data $\alpha_\text{lit}(\lambda)$ comprises two datasets: 
(1) the absorbance data in the range \SI{228}{nm}--\SI{250}{nm} \cite{Sugita1971circular}, which we refer to
as ``ultraviolet'' (UV)
and (2) the data in the range \SI{250}{nm}--\SI{1100}{nm} \cite{Friebel2005concentratedHb}, referred to here as ``visible'' (VIS). Although  our use of the terms ``visible'' and ``ultraviolet''
deviates from the usual definition, it is convenient to distinguish between the two datasets:
\begin{equation}
 \alpha_\text{lit}(\lambda) = \begin{cases}
                               \alpha_\text{UV}(\lambda) = \text{data from \cite{Sugita1971circular}} & \lambda \in [228, 250]\,\si{nm},\\
                               \alpha_\text{VIS}(\lambda) = \text{data from \cite{Friebel2005concentratedHb}} & \lambda \in [250, 1100]\,\si{nm},\\
                               0 & \text{else}.
                              \end{cases}
\end{equation}
The numerical KK transform is applied to both parts separately, such that
$G_\text{lit}(\lambda) = G_\text{UV}(\lambda) + G_\text{VIS}(\lambda)$.

\subsubsection{Fitting of Free Parameters}
Neither of the two free parameters of the model, $a_L$ and $a_\delta$ can be computed from  literature data a priori with satisfying accuracy. 
For the peptide absorption $a_L$, the order of magnitude can be estimated from the semi-quantitative data \cite{Woods1970absorptionproteins},
where the absorbance maximum is given.
It is important to keep in mind that the KK transform of the peptide-peak in the deep UV depends much stronger on the center position and the area under the peak
than on its actual maximum. Since the peak shape is not quantitatively known, the peak height does not contain enough information to determine $a_L$.

The real RI 
\begin{equation}
 n(\lambda; a_L, a_\delta) = n_{\rm H_2O}(\lambda) + c_{\rm Hb}\, B(\lambda; a_L, a_\delta)
\end{equation}
and the real refractive increment 
\begin{equation}
 B(\lambda; a_L, a_\delta) = G_\text{lit}(\lambda) \underbrace{- \frac{n_{\rm H_2O}(\lambda) - 1}{\rho_{\rm{Hb}}}}_{=: G_{\rm H_2O}(\lambda)}   + a_L\,\widetilde{G}_L(\lambda) +  a_\delta \label{eq:B_linear_parameters}
\end{equation}
linearly depend on the parameters $a_L$ and $a_\delta$. Thus we use a linear least squares approach to
optimize the parameter values.

The empirical model function in \cite{Friebel2006modelfunction} is formally identical to ours,
but the quantity $\beta(\lambda) = B(\lambda)/n_{\rm H_2O}(\lambda)$
was considered instead of $B(\lambda)$, cf. \eqref{eq:FriebelModel} and  \eqref{eq:realRI_linear}.
The measurements   are given at wavelengths $\lambda_i$, $i=1,\dotsc, N$
and we denote these experimental values by $\beta^*(\lambda_i)$. We convert them to
$B^*_i = 
\beta^*(\lambda_i)\,n_{\rm H_2O}(\lambda_i)$. In the
following, the $B^*_i$
will be referred to simply as  ``the measurement data''.

At each wavelength, the increment $B_i:=B(\lambda_i)$ consists of a fixed part
resulting from numerical KK transformation
\begin{equation}
  {B}_{0, i} = {G}_\text{lit}(\lambda_i) + {G}_{\rm H_2O}(\lambda_i) \label{eq:constant_contrib_B}
\end{equation}
and a function yet to be determined, which models the deep UV contributions
\begin{equation}
 f_i = a_L\widetilde{G}(\lambda_i) + a_\delta  { \stackrel{\text{def. } h_r(\lambda)}{=} \sum_{r=L,\delta}a_r\,h_r(\lambda_i).}
\end{equation}
Switching to matrix-vector notation, we write this function vector as
$\bm{f} = \mathsf{H}\,\bm{a}$, with $\mathsf{H} := \{h_r(\lambda_i)\}_{i r}\in\mathbb{R}^{N\times 2}$
and $\bm{a} = (a_L, a_\delta)^T$ is the parameter vector.

Now we want to minimize the deviation between KK results with deep UV absorbance model $\bm{B}$ and measurement data $\bm{B}^*$
\begin{equation}
 \bm{B}^* - \bm{B} = \bm{y} - \bm{f}, 
\end{equation}
where $\bm{y}  = \bm{B}^* -  \bm{B}_0$ is the data vector.  The entries of vector $\bm{B}_0$ are given in \eqref{eq:constant_contrib_B}.
The linear least squares problem  is then  $\chi^2(\bm{a}) \to \min$ with
\begin{align}
 \chi^2(\bm{a}) :={}& (\bm{y} - \bm{f})^T\,\mathsf{W}\,(\bm{y}-\bm{f}) \\
 ={}& \sum_{\lambda_i =\SI{250}{nm}}^{\SI{1100}{nm}} w_{ij} [B^*(\lambda_i)-B(\lambda_i; \bm{a})]\,[B^*(\lambda_j)-B(\lambda_j; \bm{a})],\nonumber
\end{align}
where $\mathsf{W} = \{w_{ij}\}_{i,j=1}^N$ is a weight matrix given by the inverse of the covariance matrix $\mathsf{V}$ of the data vector.
The conditions for minimal   $\chi^2(\bm{a})$  are solved by standard linear algebra, which yields
\begin{align}
  \hat{\bm{a}} = \arg\min\chi^2(\bm{a}) &= {(\mathsf{H}^T\mathsf{V}^{-1}\mathsf{H})^{-1}\mathsf{H}^T\mathsf{V}^{-1}}\bm{y}\\
 \hat{\bm{f}} = \mathsf{H}\,\hat{\bm{a}} &= \underbrace{\mathsf{H}\,(\mathsf{H}^T\mathsf{V}^{-1}\mathsf{H})^{-1}\mathsf{H}^T\mathsf{V}^{-1}}_{=:\mathsf{F}}\bm{y}
 \label{eq:fit_fFy}
\end{align}
for parameter and function vector and $\hat{\bm{B}} = \bm{B}_0 + \hat{\bm{f}}$ for the refractive increment.
The resulting amplitude of the Lorentzian is shown in Fig.~\ref{fig:spectra} as generic peptide absorbance.

\subsubsection{Deoxyhemoglobin}
Both, the absorption spectra of oxygenated and deoxygenated hemoglobin are well known, cf. Fig.~\ref{fig:spectra}.
However, the method of \cite{Friebel2005concentratedHb} to measure the real RI
could only be applied to oxyhemoglobin, whereas measurements with deoxyhemoglobin were
not possible due to  precipitation resulting in backscattering of light from highly  concentrated solutions
of deoxyhemoglobin.
Nevertheless, the KK analysis allows to derive 
a result for the real refractive increment of deoxyhemoglobin. To this end, we use the same model for the deep UV absorbance as above, Lorentzian and delta-peak,
with the coefficients $\hat{a}_L, \hat{a}_\delta$ found  by the least-squares fit for oxyhemoglobin.
Combining this with the KK transform of the spectrum $\alpha^\text{deoxy}(\lambda)$ for deoxyhemoglobin, we calculate the real refractive increment according to
\begin{align}
 \begin{split}
 B^\text{deoxy}(\lambda_i) &= B_0^\text{deoxy}(\lambda_i) + \hat{f}(\lambda_i)\\
 &= G_\text{lit}^\text{deoxy}(\lambda_i) + G_{\rm H_2O}(\lambda_i) + \hat{f}(\lambda_i). \label{eq:Bdeoxy}
 \end{split}
\end{align}
% The result is shown in Fig.~\ref{fig:B_fit_deoxy} together with the results for oxyhemoglobin.
The reason to simply use the same deep UV model curve for oxyhemoglobin as well as for deoxyhemoglobin is  that oxygenation affects the heme-groups in the hemoglobin
complex. The absorption peaks at about \SI{420}{nm} and \SI{560}{nm} are altered by oxygenation  (cf. Fig.~\ref{fig:spectra}) 
but not the peptide backbone causing the deep UV absorption.

\subsection{Uncertainties}
Both, the Kramers-Kronig transformation and the linear least-squares fit to the data 
are linear transformations, which can formally be carried out by matrix multiplication.
Hence, it is easy to perform the uncertainty propagation in terms of mean values and covariance matrices.
A detailed description is given in appendix~\ref{app:uncertainties}.
All values for uncertainties given here correspond to one standard deviation.

An important uncertainty contribution  stems from the uncertainty of the hemoglobin concentration $c_{\rm Hb}$.
Because this influences the concentration-specific transmittance and reflectance spectra by a global factor,
the uncertainties of the resulting real refractive increment are strongly correlated
between all wavelengths, \ie, the correlation coefficient
\begin{equation}
 r_{ij}: = \frac{\cov(B_i, B_j)}{\sqrt{\var(B_i)\, \var(B_j)}}
\end{equation}
is close to $+1$ even if $\lambda_i-\lambda_j$ is large. Here ``$\cov$'' and ``$\var$'' denote covariance and variance, respectively.
In such a case the use  of  the diagonal elements of the covariance matrix  only, \ie, the variances, may lead to
unnecessarily less significant results with respect to the uncertainty estimation.
When, for instance, the difference between two random variables $X,Y$ is considered, the variance is
\begin{equation}
 \var(X-Y) = \var(X) + \var(Y) - 2\,\cov(X,Y),
\end{equation}
which may be small even if $\var(X),\var(Y)$ are large.

\section{Results and Discussion}\label{sec:results}
\begin{figure}[t]
 \centering
 \includegraphics[scale=0.80]{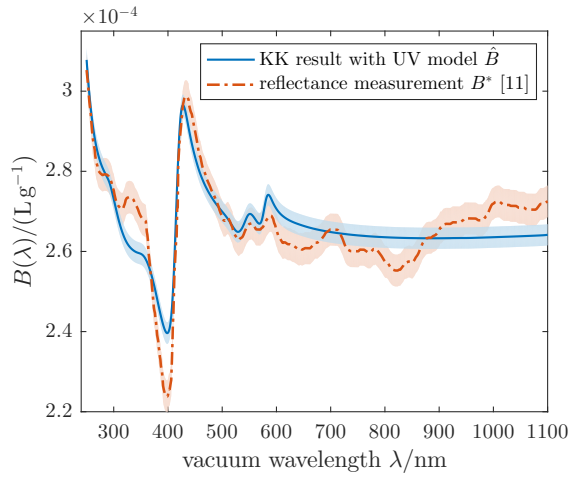}
 \caption{
  Real refractive increment $B(\lambda)$ of aqueous solutions of  oxyhemoglobin: Experimental data \cite{Friebel2006modelfunction} and 
  results from this work fitted to the data. The shaded bands indicate the  measurement uncertainties given in \cite{Friebel2006modelfunction}
  and computed in appendix~\ref{app:uncertainties}, respectively.}
 \label{fig:B_fit}
\end{figure}

\begin{figure}[t]
 \centering
 \includegraphics[scale=0.80]{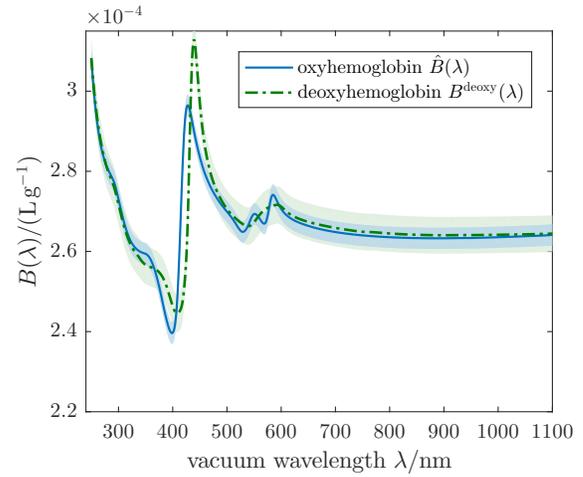}
 \caption{
  Real refractive increment of aqueous hemoglobin solutions: Result for deoxyhemoglobin obtained by KK relations 
  with the same model for deep UV absorbance as for oxyhemoglobin. 
  Cf. Fig.~\ref{fig:B_fit}. Estimated standard deviations (shaded bands) are larger for deoxyhemoglobin because $B^\text{deoxy}(\lambda)$ is composed of more
  terms with uncorrelated concentration uncertainties than $\hat{B}(\lambda)$. 
  }
 \label{fig:B_fit_deoxy}
\end{figure}

\begin{figure}[t]
 \centering
 \includegraphics[scale=0.80]{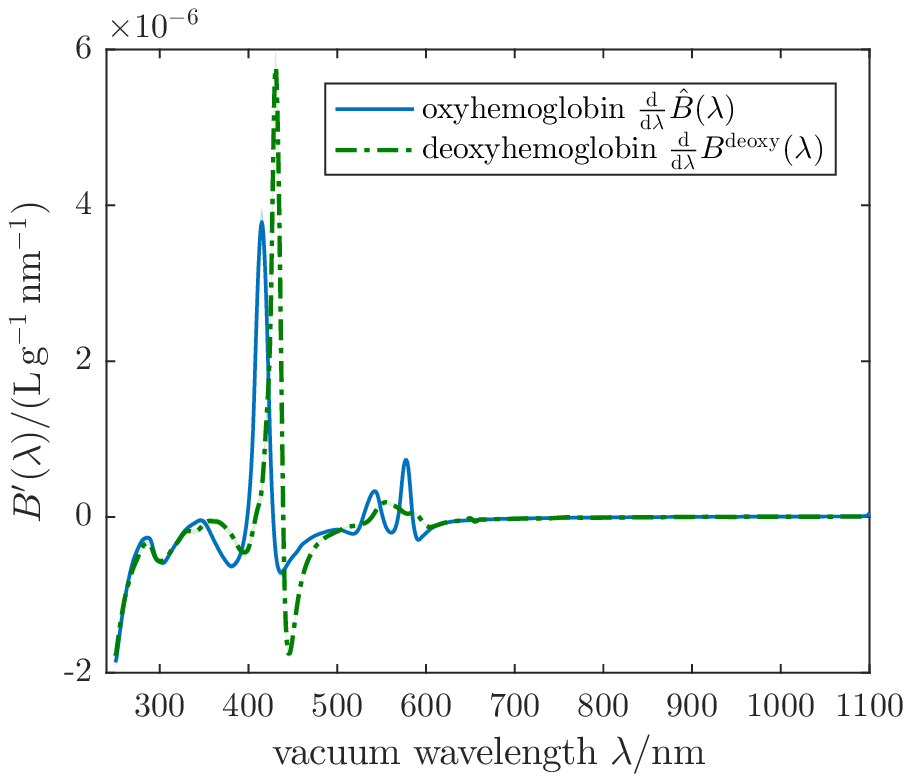}
 \caption{
  Derivative of the real refractive increment $B'(\lambda) = \frac{\dif}{\dif \lambda}B(\lambda)$: Comparison between results for deoxyhemoglobin and oxyhemoglobin. 
  Estimated standard deviations (shaded bands, barely visible) of $B'(\lambda)$ are much smaller than for $B(\lambda)$ (Fig.~\ref{fig:B_fit_deoxy}). 
  This is because most of the (strongly correlated) concentration uncertainty cancels out  in the derivative.}
 \label{fig:B_prime}
\end{figure}

\subsection{Real Part of Refractive Index for Oxy- and Deoxyhemoglobin}
Fig.~\ref{fig:B_fit} shows the result of the presented KK analysis for the real refractive increment $\hat{B}(\lambda)$ for oxyhemoglobin along with the experimental data $B^*(\lambda)$ from \cite{Friebel2006modelfunction}.
The uncertainties are shown as shaded bands, where the half-width  of the band corresponds to one  standard deviation derived by covariance-matrix calculus.
The analogue result for deoxygenated hemoglobin is shown in Fig.~\ref{fig:B_fit_deoxy}.
Fig.~\ref{fig:B_prime} shows the derivative $\hat{B}'(\lambda) = \dif\hat{B}(\lambda)/\dif\lambda$. Here the
strong correlation is eliminated, resulting in much smaller uncertainties. 
It is evident from Fig.~\ref{fig:B_prime} that the main uncertainty of the result
concerns the exact vertical position of the curve $\hat{B}(\lambda)$, whereas the uncertainties related to its shape, described equivalently by the derivative $\hat{B}'(\lambda)$, are smaller.

% \subsection{Discrepancies Between Results and Literature Values}
The result $\hat{B}(\lambda)$ of the KK analysis for oxyhemoglobin  and  the experimental data $B^*(\lambda)$
as  shown in Fig.~\ref{fig:B_fit} have a similar overall shape, but there are deviations which 
cannot be accounted for by the measurement uncertainties of $B^*(\lambda)$ given in \cite{Friebel2006modelfunction}
and the uncertainties of the KK result derived as described in appendix~\ref{app:uncertainties}.
For instance, the difference between the peak at 433\,nm and the dip at 399\,nm for $B^*$ is
% peak-peak amplitudes between 433nm and 399nm:
% B*:    (7.4945 +/- 0.5695)*10^-5 L/g
% B_fit: (5.4670 +/- 0.0575)*10^-5 L/g
\[
%  [B^*]^{\SI{433}{nm}}_{\SI{399}{nm}} = (7.5\pm0.6)\times 10^{-5}\,\si{L.g^{-1}}
 B^*(\SI{433}{nm}) - B^*(\SI{399}{nm}) = (7.5\pm0.6)\times 10^{-5}\,\si{L.g^{-1}}
\]
while that for our result is
\[
%  [\hat{B}]^{\SI{433}{nm}}_{\SI{399}{nm}} = (5.47\pm0.06)\times 10^{-5}\,\si{L.g^{-1} }.
\hat{B}(\SI{433}{nm}) - \hat{B}(\SI{399}{nm})= (5.47\pm0.06)\times 10^{-5}\,\si{L.g^{-1} }.
\]
The small  value for the  uncertainty of $\hat{B}(\SI{433}{nm}) - \hat{B}(\SI{399}{nm})$ is due to the fact that the strongly correlated
concentration contributions to the uncertainty cancel out almost  completely for this difference. This observation also holds  for the derivative $\hat{B}'(\lambda)$ (cf. Fig.~\ref{fig:B_prime}).
The difference between the two values clearly exceeds the uncertainties.

The deviations between $\hat{B}(\lambda)$ and $B^*(\lambda)$ can not be attributed to unknown spectral absorptions
outside the  \SI{250}{nm}--\SI{1100}{nm} range, since features producing % (after KK transformation)
such discrepancies would  necessarily be inside this wavelength range.
We have modeled important spectral absorptions at the UV end of the spectrum. Concerning possible IR absorptions that 
are not considered in the proposed model, we give the following notes:
The absorption spectra of aqueous hemoglobin solutions in the IR between \SI{1.1}{\micro m} and \SI{2.6}{\micro m} are dominated by water \cite{Friebel2005concentratedHb}
indicating that the imaginary refractive increment of Hb is very low in this region. Hypothetical absorption lines due to Hb at even longer wavelengths would contribute to the real refractive index
increment $B(\lambda)$ in the form of constants or functions with a gentle negative slope. Any possible influence of the long-wavelength end of the spectrum of Hb
would thus change the agreement between our result $\hat{B}(\lambda)$ and the literature data $B^*(\lambda)$ for the worse. Thus we conclude that no
important contribution to the absorption spectrum was missed at the long-wavelength end.
The KK relations themselves are valid as long as the framework within which the data were measured holds, \ie,
 classical electrodynamics and linear, causal media.
Optical activity of Hb has been examined previously \cite{Sugita1971circular, Laasberg1971optical} and can be ruled out as well, since the expected effect is too weak.
Furthermore, we can  exclude  biological variability,  since absorption and reflectance/refraction data  were measured on the same type of sample 
\cite{Friebel2005concentratedHb, Friebel2006modelfunction}.
This implies that the absorption and refraction data presented in \cite{Friebel2005concentratedHb, Friebel2006modelfunction}
are not self-consistent with the given measurement uncertainties.

Lastly, we address the question, whether these discrepancies are of practical importance.
At $\lambda=\SI{399}{nm}$, where the discrepancy is largest, one has 
${B}^* = \SI{2.24e-04}{L.g^{-1}}$ from \cite{Friebel2006modelfunction} and  $\hat{B} = \SI{2.41e-04}{L.g^{-1}}$ for our analysis.
At a concentration of $c_{\rm{Hb}}=\SI{340}{g.L^{-1}}$ the resulting real RIs are
$n^* = n_{\rm H_2O} + c_{\rm Hb} \,B^*= 1.420$ and $\hat{n} = n_{\rm H_2O} + c_{\rm Hb} \,\hat{B} = 1.426$, respectively and the residual is
$|\hat{n}-n^*| = 0.006 = 0.4\%\,n^*$. Since this value is in the sub-percent region
it may, at first glance, seem unimportant whether one uses $n^*$ or $\hat{n}$.
However, when RBCs are examined with optical methods such as phase contrast microscopy or
light scattering in a flow cytometer, they are usually immersed in water with an RI of $n_{\rm H_2O}(\SI{399}{nm}) = 1.344$,
or a saline solution of slightly higher RI.
Hence the complex \emph{contrast} between cell and surrounding medium is 
\begin{equation}
 \mathfrak{m}-1 := \frac{\mathfrak{n}}{n_{\rm H_2O}}-1 = \begin{cases}
                                     \num{0.061+0.008i} & \text{\cite{Friebel2006modelfunction}}\\
                                    \num{0.057+0.008i} & \text{our result}
                                  \end{cases}.
\end{equation}
In phase contrast microscopy, the signal for the optical thickness of the cell is directly proportional to the contrast $\mathfrak{m}-1$.
It is also a parameter governing the light scattering by cells.
The Rayleigh-Gans-Debye theory is an approximate description of light scattering for the limiting case of particles with low contrast and moderate size.
It predicts a scattered electric field proportional to $\mathfrak{m}-1$.
The real parts of the above two values for $\mathfrak{m}-1$ differ by more than 7\%, indicating that any quantitative analysis 
of the interaction of RBCs with light that requires a priori knowledge of the contrast can not be more accurate than this.%
%as long as it is not clear what causes the discrepancy.

\subsection{Comparison to Previous KK Analyses}\label{subsec:comparison_previous_KK}
\begin{figure}[t]
  \centering
  \includegraphics[scale=0.80]{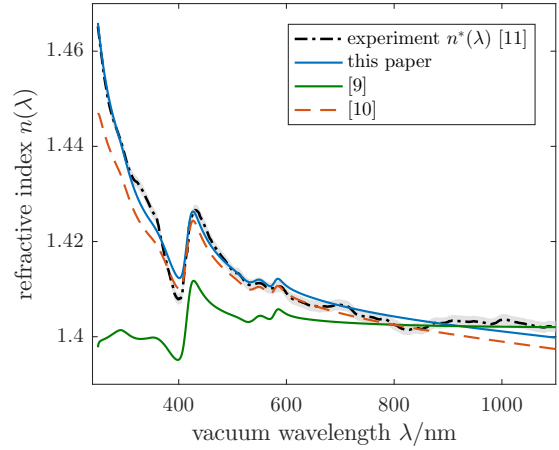}
  \caption{
%    Comparison of different methods to calculate the real RI of Hb solutions for oxyhemoglobin at a concentration of $c_{\rm Hb}= \SI{287}{g.L^{-1}}$:
   Comparison between the experimental real RI derived from measured refractive increment of Hb solutions  \cite{Friebel2006modelfunction} and the water RI  \cite{Daimon2007refractivewater}
   (measurement uncertainty indicated by shaded band)
   and different Kramers-Kronig analyses. Concentration is $c_{\rm Hb}= \SI{287}{g.L^{-1}}$.
   The method in \cite{faber2004oxygen} ignores absorption outside the measured range and
   in \cite{Sydoruk2012refractiveindex}
   the UV and IR absorption of water was taken into account. In  the present work
   the excluded water fraction was accounted for, as well as a model for deep UV absorbance.
   All KK methods were applied to the same dataset depicted in Fig.~\ref{fig:spectra}. 
   Curves calculated according to \cite{faber2004oxygen,Sydoruk2012refractiveindex}
   are matched to the experimental curve at $\lambda_0=\SI{800}{nm}$. The two free parameters characterizing the deep UV model in our method are
   determined by a global fit to the experimental
   refractive increment. 
 }
 \label{fig:literature_KK_methods}
\end{figure}
We will now briefly review the previous investigations employing KK relations \cite{faber2004oxygen, Sydoruk2012refractiveindex} and compare the methods.
In \cite{faber2004oxygen}, the authors started from \eqref{eq:KK_nRe} and applied it to the Hb absorption spectra in a finite
spectral range, \ie, instead of \eqref{eq:ansatz_absorption_2} they assumed
\begin{equation}
 \kappa_\text{\cite{faber2004oxygen}}(\lambda) =  c_{\rm Hb}\,\alpha_\text{VIS}(\lambda) = 
				  c_{\rm Hb}\begin{cases}
                                  \,\alpha(\lambda) & \lambda\in[a,b]  \\
                                  0 & \text{else}
                                 \end{cases},
\end{equation}
where $[a,b]$ is the measured spectral range, \ie, here  $[a,b] = [250, 1100]\,\si{nm}$.
The authors then used a subtractive form of the KK relations, where the difference $n(\lambda)-n(\lambda_0)$ is considered
which yields
\begin{equation}
 n_\text{\cite{faber2004oxygen}}(\lambda) =  n(\lambda_0)  + c_{\rm Hb}\,[ G_\text{VIS}(\lambda) - G_\text{VIS}(\lambda_0)]. \label{eq:nFaber}
\end{equation}
Here, $ G_\text{VIS}(\lambda):=\KK[\alpha_\text{VIS}](\lambda)$ is the dispersion resulting from the measured spectrum.
The free parameter $n(\lambda_0)$ is fixed by a refractometric measurement at wavelength $\lambda_0=\SI{800}{nm}$.
If the non-subtractive KK relations had been used instead, the result would have been
\begin{equation}
 n(\lambda) = 1+  c_{\rm Hb}\,G_\text{VIS}(\lambda),
\end{equation}
which is off the true value by a significant amount. One can  remove this discrepancy  by replacing the  1
in the above expression by a free parameter, which can be interpreted as  
 deep UV absorption. 
Again, this free parameter can  be fixed by a single measurement at $\lambda_0$. The result is then the same as in \eqref{eq:nFaber}.
However,  the subtractive KK transform $\KK[\alpha](\lambda) -\KK[\alpha](\lambda_0)$ can be re-written into a single integral where the kernel decays faster than in the standard KK relations,
which is  numerically favorable and thus given as an argument for the use of subtractive relations.
Thus,  the subtractive form of KK relations masks the lack of knowledge of
the absorption spectrum outside the measured spectral range:
At least one important absorption peak at short wavelengths has apparently been omitted.
As long as the missing peak is  far away from the region of interest, the simple
addition of a constant works fine. However,
if the location of the peak becomes important as is the case for water, this model is insufficient.

In \cite{Sydoruk2012refractiveindex}, the authors made the ansatz
\begin{equation}
 \kappa_\text{\cite{Sydoruk2012refractiveindex}}(\lambda) = \kappa_{\rm{H_2O}}(\lambda) + c_{\rm Hb}\,\alpha_\text{VIS}(\lambda).
\end{equation}
for the imaginary RI of the Hb solution, which takes the absorption due to water into account but neglects the excluded volume effect for water
in contrast to our approach
in \eqref{eq:ansatz_absorption_2}. Apart from this difference, the formal
application of the KK relations in the present work is identical to  that in \cite{Sydoruk2012refractiveindex}. The result was
\begin{equation}
 n_\text{\cite{Sydoruk2012refractiveindex}}(\lambda) =  n_{\rm H_2O}(\lambda)  + c_{\rm Hb}\, G_\text{VIS}(\lambda),
\end{equation}
which provides also a theoretical derivation of the empirical finding $n(\lambda) = n_{\rm H_2O}(\lambda)[1  + c_{\rm Hb}\,\beta(\lambda)]$ reported in \cite{Friebel2006modelfunction}.
However, the result that $\beta(\lambda) = G_\text{VIS}(\lambda)/n_{\rm H_2O}(\lambda)$ is incomplete, as we have discussed.
Subtractive KK relations were used as well to match the RI at $\lambda_0=\SI{800}{nm}$.

\begin{figure}[t]
  \centering
  \includegraphics[scale=0.80]{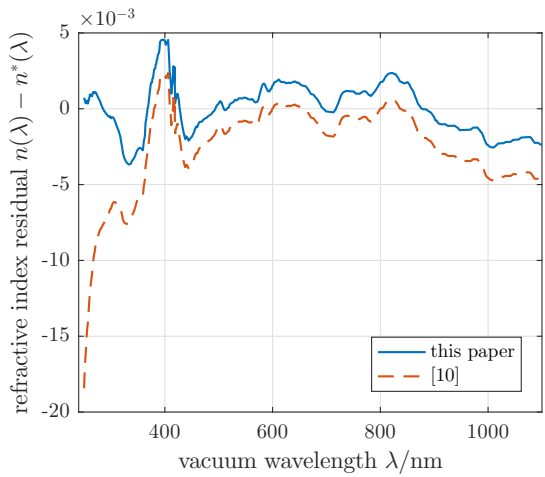}
  \caption{
    Residuals   between calculated real RI $n(\lambda)$ and experimental real RI $n^*(\lambda)$  \cite{Friebel2006modelfunction}
    for our method and the method applied in \cite{Sydoruk2012refractiveindex}, cf. Fig.~\ref{fig:literature_KK_methods}.
    }
    \label{fig:literature_KK_methods_residuals}
\end{figure}
To compare these two previously presented methods with ours, we applied them to the spectra of oxyhemoglobin presented in \cite{Friebel2005concentratedHb} and shown in Fig.~\ref{fig:spectra}. 
As an example, a concentration of $c_{\rm Hb} = \SI{287}{g.L^{-1}}$ was assumed and $n(\lambda_0 = \SI{800}{nm}) = 1.403$ was taken from \cite{Friebel2005concentratedHb}
as well. 
The comparison of the two methods applied in \cite{faber2004oxygen, Sydoruk2012refractiveindex} with  our method
and the experimental dispersion curve given in \cite{Friebel2006modelfunction}
is shown in Fig.~\ref{fig:literature_KK_methods}.
Neglecting the water influence as in \cite{faber2004oxygen} yields a dispersion curve which
substantially deviates from the measurement (dash-dotted black line in Fig.~\ref{fig:literature_KK_methods}) everywhere,
except at and above $\lambda_0$, where it was matched.
With the water absorption and the resulting dispersion taken into account \cite{Sydoruk2012refractiveindex} the agreement with the experimental
result is already much better. However, the result from  \cite{Sydoruk2012refractiveindex}  and experimental curve substantially differ in the
UV below \SI{375}{nm}. The agreement in the UV is much better in our approach.
This is also evident from Fig.~\ref{fig:literature_KK_methods_residuals}, where the residuals between KK computations and experiment are plotted.
% Note, however that in all three models (cf. Fig.~\ref{fig:literature_KK_methods}), the
% difference between the peak at \SI{433}{nm} and the valley at \SI{399}{nm} is more pronounced in the
% measured RI curves than in the curves computed from experimentally measured spectra with KK relations.

\section{Summary}\label{sec:summary_outlook}
The complex refractive index of a hemoglobin  solution, which forms the cytoplasm of erythrocytes and determines their optical properties can be computed as
\begin{equation}
 \mathfrak{n}(\lambda) = n_{\rm H_2O}(\lambda)+ \imag\,\kappa_{\rm H_2O}(\lambda)  + c_{\rm Hb}[B(\lambda) + \imag\, \alpha(\lambda) ],
\end{equation}
where $\kappa_{\rm H_2O}(\lambda)$ is negligibly small for $\lambda\in[250,1100]\,\si{nm}$ and physiological hemoglobin concentrations $c_{\rm Hb}$.
In the present work, we have computed the real refractive increment $B(\lambda)$ from experimental spectra of the imaginary
refractive increment $\alpha(\lambda)$ for $\lambda\in[250,1100]\,\si{nm}$.
We formally separated the solution's imaginary RI into a water and a hemoglobin part and then applied the Kramers-Kronig relations to obtain the
real RI and thus an expression for $B(\lambda)$, \eqref{eq:realRI_linear}.
The absorption spectra available in the literature  \cite{Sugita1971circular, Friebel2005concentratedHb} do not resolve the strong UV absorbance of hemoglobin's peptide-backbone. Hence we modeled
it by a Lorentzian line of unknown amplitude, which introduces a free parameter $a_L$ into the expression for $B(\lambda)$. A second free parameter $a_\delta$ is introduced
as a wavelength-independent term accounting for extreme UV absorbance, cf. \eqref{eq:B_linear_parameters}. These two free parameters were determined by a linear 
least squares fit to the literature data $B^*(\lambda)$ \cite{Friebel2006modelfunction}, the result of the fit is denoted by $\hat{B}(\lambda)$.

We evaluated spectra for oxygenated and deoxygenated hemoglobin.
Data files  of the results are provided as supplementary material for $\hat{B}(\lambda)$ (Data File 1), ${B}^\text{deoxy}(\lambda)$ (Data File 3), the corresponding covariance matrices (Data Files 2 and 4), 
the converted literature data for $\alpha(\lambda)$ and $\alpha^\text{deoxy}(\lambda)$ \cite{Friebel2005concentratedHb}
and the values for $n_{\rm H_2O}(\lambda)$ computed according to \cite{Daimon2007refractivewater}
 (Data Files 1 and 3) as well as for $\hat{B}'(\lambda)$ (Data File 5).

The uncertainties for the curve $\hat{B}(\lambda)$ were computed and reveal that its shape is resolved much more accurately than in the measurements $B^*(\lambda)$ in \cite{Friebel2006modelfunction}.
The  analysis furthermore shows that the real refractive increments for  oxygenated and deoxygenated hemoglobin differ significantly from each other for wavelengths between \si{350}{nm} and
\si{600}{nm}.   
In the vicinity of the absorption band at \si{420}{nm}, deviations between our result for the real refractive increment and the experimental values \cite{Friebel2006modelfunction}
clearly exceed the measurement uncertainties, where the curve obtained by our KK analysis has a much smoother shape and does not exhibit the 
non-monotonic up-and-down movements for wavelengths larger than 600 nm  of the data presented in \cite{Friebel2006modelfunction}.

%%%%%%%%%%%%%%%%%%%%%%%%%%%%%%%%%%%%%%%%%%%%%%%% APPENDIX %%%%%%%%%%%%%%%%%%%%%%%%%%%%%%%%%%%%%%%%%%%%%%%%%%%%%%%%%%%%%%%%
% \begin{strip}
% \hrule
% \end{strip}

\appendix
%\part*{Appendix}
\section{Appendix: Numerical Integration Scheme for Kramers-Kronig Relations}\label{app:num_int}
For numerical integration of KK relations, we follow the concept described in \cite{Emeis1967numericalKK}: a Riemann sum with Taylor expansion at the singularities of the integrand.
Numerical stability was tested by comparing to the analytical transformations of different Lorentzian and rectangular profiles.

We want to evaluate numerically the expression
\begin{align}
 \begin{split}
 \pi\,G(\lambda) &= -2 \dashint_{\lambda_a}^{\lambda_b} \frac{\lambda}{\Lambda} \frac{\lambda}{\Lambda^2-\lambda^2}  \,\alpha(\Lambda)\, \dif \Lambda \\
  &= \dashint_{\lambda_a}^{\lambda_b} \left( \frac{2}{\Lambda}  - \frac{1}{\Lambda+\lambda} -  \frac{1}{\Lambda-\lambda}\right)  \,\alpha(\Lambda)\, \dif \Lambda.
 \end{split}
\end{align}
The measurement data are given at discrete points
\begin{align}
 \alpha_i &:= \alpha(\lambda_i),\\
 \lambda_i &:= \lambda_a + t\left(i-\frac12\right), \qquad i=1, \dotsc, N,
\end{align}
where $\lambda_1 = \SI{250}{nm}$, $\lambda_N = \SI{1100}{nm}$ and $t=\SI{1}{nm}$. Hence  $\lambda_a = \SI{249.5}{nm}, \lambda_b = \SI{1100.5}{nm}$ and $N= 851$.
We only evaluate the integral at the grid points $G_i:=G(\lambda_i)$.
The third term in the integral has a singularity at $\lambda = \Lambda$.
The first two terms are not singular, hence no principal value integrals have to be used here.
All integrals for non-singular integrands are approximated by Riemann sums, including the third term for $\Lambda\notin [\lambda_i-t/2, \lambda_i+t/2]$.
The remaining principal value integral can be re-written by Taylor series expansion of the integrand. 
Using only the lowest non-vanishing order yields
\begin{equation}
 \pi\,G_i \approx  \sum_{j=1}^N\left( \frac{2}{\lambda_j}- \frac{1}{\lambda_j + \lambda_i}\right)\,\alpha_j\,t   
 - \sum_{\stackrel{ j\neq i}{j=1}}^N \frac{1}{\lambda_j - \lambda_i}\,\alpha_j\,t - t\,\alpha'_i. \label{eq:num_KK}
\end{equation}
Numerically, we use the nearest-neighbor lattice-derivatives
\begin{align}
 t\,\alpha'_i = \begin{cases}
                 ({\alpha_{i+1} - \alpha_{i-1}})/{2} & 1<i<N\\
                 \alpha_2 - \alpha_1 & i=1\\
                 \alpha_N - \alpha_{N-1}& i=N
                \end{cases}.
\end{align}
% \begin{align}
%  t\,\alpha'_i &= \frac{\alpha_{i+1} - \alpha_{i-1}}{2}\quad\text{ for }\quad 1<i<N, \\
%   t\,\alpha'_1 &= \alpha_2 - \alpha_1, \\
%  t\,\alpha'_N &= \alpha_N - \alpha_{N-1}.
% \end{align}
Note that \eqref{eq:num_KK} can also be written as $\bm{G} = \mathsf{K}\,\bm{\alpha}$, where $\mathsf{K}$ is a  $N\times N$ matrix.

\section{Appendix: UV Domain Kramers-Kronig-Transform of Lorentzian Curve}\label{app:trafo_Lorentzian}
If the KK transform (\eqref{eq:KK_nRe}  is applied to
the antisymmetric Lorentzian line in the wavelength domain \eqref{eq:antisymm_Lorentzian}
the result is  
\begin{align}
\begin{split}
 \KK[ \alpha_L] (\lambda) 
  ={}& a_L\frac1\pi \left(   \frac{\lambda-L}{(\lambda-L)^2+\Gamma^2} +  \frac{L}{L^2+\Gamma^2} \right)
  \\
  &- 
    a_L\frac1\pi \left(   \frac{\lambda+L}{(\lambda+L)^2+\Gamma^2} -  \frac{L}{L^2+\Gamma^2} \right).
\end{split}
\end{align}
In our work, however, we need to integrate only over the deep ultraviolet spectrum below a threshold $\lambda_a$.
The analytical expression for this reads
\begin{align}
\begin{split}
  G_L(\lambda)
  ={}&
  \frac{1}{\pi} \dashint_{-\lambda_a}^{\lambda_a}\left(\frac{1}{\Lambda} - \frac{1}{\Lambda-\lambda}\right)\,\alpha_L(\Lambda)\,\dif \Lambda
  \\
  &
  = \frac{a_L}{\pi^2} \bigg\{
    \frac12 \ln\left(\frac{ (\lambda_a-L)^2 + \Gamma^2 }{ (\lambda_a+L)^2 + \Gamma^2 }\right)
  \\
  &
  \times
    \left[  \frac{\Gamma}{ (\lambda-L)^2 + \Gamma^2 } + \frac{\Gamma}{ (\lambda+L)^2 + \Gamma^2 }  -  \frac{2\Gamma}{ L^2 + \Gamma^2 }  \right]
    \\
   &-  
    \ln\left(\left|\frac{\lambda_a-\lambda}{\lambda_a+\lambda}\right|\right) \times
    \left[  \frac{\Gamma}{ (\lambda-L)^2 + \Gamma^2 } - \frac{\Gamma}{ (\lambda+L)^2 + \Gamma^2 }  \right]	
    \label{eq:partialUVKKLorentz}
    \\
  & +
    \left( \pi - \arctan\frac{\Gamma}{\lambda_a-L} - \arctan\frac{\Gamma}{\lambda_a+L} \right)
  \\
  &
  \times 
    \left[  \frac{\lambda-L}{ (\lambda-L)^2 + \Gamma^2 } - \frac{\lambda+L}{ (\lambda+L)^2 + \Gamma^2 }  +  \frac{2L}{ L^2 + \Gamma^2 }  \right]
  \bigg\} 
  \end{split}
\end{align}
for $\lambda>\lambda_a$.
\eqref{eq:partialUVKKLorentz} is useful to describe the deep UV part of the absorption spectrum without the need for numerical integration.

We also define the transformation for unit amplitude $\widetilde{G}_L(\lambda)$, such that
 $G_L(\lambda) = a_L\,\widetilde{G}_L(\lambda)$.

\section{Appendix: Details of Uncertainty Propagation}\label{app:uncertainties}
\subsection{Uncertainties of Input Data}
We now recapitulate how spectral data are measured, in order
understand which  contributions to the combined uncertainties 
occur in the problem at hand.

The absorption spectra (inverse absorption length $\mu_a(\lambda)$, imaginary RI $\kappa(\lambda)$)
are measured via the attenuation of a beam of light passing through a sample of known thickness.
This is performed for a number of wavelengths $\lambda_i$, $i=1,\ldots, N$.
The molar extinction coefficient $\varepsilon_M(\lambda)$ and the imaginary refractive increment  $\alpha(\lambda)$ 
are obtained from these quantities by dividing with the separately determined molar/mass concentration of Hb in the solution.

In \cite{Friebel2005concentratedHb, Friebel2006modelfunction} 
spectral reflectance measurements were performed and evaluated using the Fresnel equation \eqref{eq:Fresnel} to obtain  the real RI $n(\lambda)$.
To determine the experimental value of the real refractive increment $B^*(\lambda)$, several curves at different concentrations $c_{\rm Hb}$ were recorded
and the slope of $n(\lambda)$ with respect to $c_{\rm Hb}$ was computed for all $\lambda$ and normalized to the known water RI to obtain $\beta^*(\lambda)$.

Having this in mind, measurement uncertainties of the following quantities need to be taken into account:
\begin{enumerate}
  \item Detector/instrument noise in the absorption spectra. This affects 
    the inverse absorption length $\mu_a(\lambda)$ in \cite{Friebel2005concentratedHb} and the molar extinction coefficient $\varepsilon_M(\lambda)$ in \cite{Sugita1971circular}.
  \item Detector/instrument noise in the reflectance spectra \cite{Friebel2005concentratedHb, Friebel2006modelfunction}, resulting in a wavelength-independent uncertainty of $\SI{3e-6}{L.g^{-1}}$ \cite{Friebel2006modelfunction} for
  the real refractive increment $\beta^*(\lambda)$.
  \item Uncertainties of the solutions'  Hb concentration $c_{\rm Hb}$.
 \item Uncertainty of the hemoglobin density $\rho_{\rm Hb}$ relating mass concentration $c_{\rm Hb}$ and volume fraction $\phi$.
  Here we assume one digit, \ie,  $u_{\rho_{\rm Hb}} = \SI{10}{g.L^{-1}} = 0.75\%\,\rho_{\rm Hb}$.
 \item Uncertainty of the complex  RI of water.  This influence is negligible.%, since the data used for the real part are accurate to at least five decimal places
%  and the imaginary part contributes only a minor correction to the measured absorption spectra.
\end{enumerate}

These uncertainties can be divided into two classes, separating 1.--2. from 3.--4.:
\begin{enumerate}%[(i)]
 \item[(i)]  Noise in the measured reflectance and transmittance spectra, affecting  the spectral data locally.
 The  corresponding quantities  are labeled with the subscript ``{noise}''. 
   As a model, we assume  white noise, \ie, two measurement errors at different wavelengths are not correlated.
 \item[(ii)]  Measurement uncertainties in scalar quantities ($c_{\rm Hb}, \rho_{\rm Hb}$) occurring as prefactors in the spectra. These affect 
   the spectra by a global factor, \ie, the errors at different wavelengths have correlation coefficient +1. The corresponding quantities are labeled with a subscript ``{conc}''.
\end{enumerate}

In addition there are model errors, which arise due to unknown absorption spectra outside the measured range and  which we discuss in appendix~\ref{sec:peak_shape}.

\subsection{Error Model and Uncertainty Propagation}

For any random vector $\bm{\xi}\in \mathbb{R}^N$ and any non-random linear transform $\mathsf{A}\in\mathbb{R}^{N\times N}$, we have for
$\bm{\eta}=\mathsf{A}\,\bm{\xi}$
\begin{align}
 \mathbb{E}(\bm{\eta}) &= \mathsf{A}\,\mathbb{E}(\bm{\xi})\\
 \mathsf{V}(\bm{\eta}) &= \mathsf{A}\,\mathsf{V}(\bm{\xi})\,\mathsf{A}^T, \label{eq:variance_propagation}
\end{align}
where $\mathbb{E}$ denotes the expectation value (or mean) and
\begin{align} 
 \left[\mathsf{V}(\bm{\xi})\right]_{ij}=
 \cov(\xi_i,\xi_j) = \mathbb{E}\left( [\xi_i-\mathbb{E}(\xi_i)]\,[\xi_j-\mathbb{E}(\xi_j)]\right)
\end{align} 
denotes the covariance matrix. This is independent of the underlying probability distribution, only implying that the first and second moments exist.
We restrict our uncertainty analysis to mean values and covariance matrix, which describes the corresponding probability distributions fully only  in the special
case of a normal distribution.

The uncertainties of the literature absorption spectra are not given; neither the noise nor the concentration uncertainties are 
quantified. We thus estimate and model them as follows.
The covariance matrix for the absorption spectra is assumed to be $\mathsf{V}(\bm{\alpha}) = \mathsf{V}_\text{noise}(\bm{\alpha}) + \mathsf{V}_\text{conc}(\bm{\alpha})$.
Here
\begin{equation}
 \mathsf{V}_\text{noise}(\bm{\alpha}) = \sigma_\text{noise,rel}^2\,\mathrm{diag}(\bm{\alpha})^2 %= 
% 				        \sigma_\text{noise,rel}^2\begin{pmatrix}
%                                                              \alpha_1^2 	& 0	& \cdots&  0\\
%                                                              0	& \alpha_2^2		& 	&	\vdots\\
%                                                              \vdots		&			&\ddots\\
%                                                              0	& \cdots 		&	&	\alpha_N^2
%                                                             \end{pmatrix}
\end{equation}
is the local/uncorrelated part due to (white) detector noise. This matrix is diagonal.
\begin{align}
 \mathsf{V}_\text{conc}(\bm{\alpha}) &= \sigma_\text{conc,rel}^2\,\bm{\alpha}\,\bm{\alpha}^T %\nonumber\\
%  &=
% 				       \sigma_\text{conc,rel}^2\begin{pmatrix}
%                                                              \alpha_1^2 	& \alpha_1\,\alpha_2	& \cdots&  \alpha_1\,\alpha_N\\
%                                                              \alpha_1\,\alpha_2	& \alpha_2^2		& 	&	\vdots\\
%                                                              \vdots		&			&\ddots\\
%                                                              \alpha_1\,\alpha_N	& \cdots 		&	&	\alpha_N^2
%                                                             \end{pmatrix},
                                                            \label{eq:variance_model_concentration}
\end{align}
is the global/correlated part due to concentration uncertainty. This matrix has a tensor-product structure.
We use ${\sigma_\text{noise,rel} =  0.5\%}$ for the relative noise level and
${\sigma_\text{conc,rel} =  1\%}$ for the relative concentration uncertainty \cite{Witt2013traceability}.

\eqref{eq:variance_model_concentration} is motivated as follows:
Concentration errors change a spectrum by a global factor, such that 
the measured value
\begin{equation}
  \bm{\alpha}_\text{meas} = (1+\xi)\bm{\alpha}, \label{eq:concentration_error}
\end{equation} 
is off the (unknown) true value $\bm{\alpha}$. We assume this error to be unbiased, \ie,
 $\mathbb{E}(\xi)=0$ and $\sqrt{\var(\xi)} =  \sigma_\text{conc,rel}$ corresponds to the relative concentration uncertainty of the solution.
\Ie, $\mathbb{E}(\bm{\alpha}_\text{meas}) = \bm{\alpha}$, and
\begin{align} 
 \mathsf{V}(\bm{\alpha}_\text{meas} ) &= \var(\xi)\,\bm{\alpha}\,\bm{\alpha}^T = \underbrace{\var(\xi)}_{=\sigma_\text{conc,rel}^2}\,\mathbb{E}(\bm{\alpha}_\text{meas})\,\mathbb{E}(\bm{\alpha}_\text{meas})^T 
 \nonumber \\
 &\approx
 \sigma_\text{conc,rel}^2\,\,\bm{\alpha}_\text{meas}\,\bm{\alpha}_\text{meas}^T,
 \label{eq:variance_concentration_error}
\end{align} 
which is \eqref{eq:variance_model_concentration}. In the following we will drop the distinction between  unknown true values and measured values and substitute the latter for the former.

This error model is formally invariant under linear transformation. 
The measured value of some quantity obtained by a linear transform $\mathsf{A}$ (\eg, the KK transform)
\begin{equation}
 \bm{\eta}_\text{meas} = \mathsf{A}\,\bm{\alpha}_\text{meas} =  (1+\xi)\mathsf{A}\,\bm{\alpha} =   (1+\xi)\,\bm{\eta},
\end{equation}
is formally identical to \eqref{eq:concentration_error} such that the variance $\mathsf{V}(\bm{\eta}_\text{meas})$ can be computed 
from $\mathbb{E}(\bm{\eta}_\text{meas})$ in the same manner as for the original quantity $\bm{\alpha}$, \ie,
$ \mathsf{V}(\bm{\eta}_\text{meas}) = \var(\xi)\bm{\eta}\, \bm{\eta}^T$.

\subsection{Uncertainty Propagation in Kramers-Kronig Relations}
For discrete data points, the KK relations can be written in matrix vector form, \ie,
$ \bm{G} = \mathsf{K}\,\bm{\alpha}$,
 where $\mathsf{K}$ is a  $N\times N$ matrix (see appendix~\ref{app:num_int}). 
If $\mathsf{V}(\bm{\alpha})$ is the (co-)variance matrix of the spectrum $\bm{\alpha}$, then
the covariance matrix of the transform $\bm{G}$ is
\begin{equation}
  \mathsf{V}(\bm{G}) = \mathsf{K} \, \mathsf{V}(\bm{\alpha}) \, \mathsf{K} ^T
\end{equation}
and with the model for $\mathsf{V}(\bm{\alpha})$ one obtains
\begin{equation}
  \mathsf{V}(\bm{G}) = \mathsf{V}_\text{noise}(\bm{G}) + \mathsf{V}_\text{conc}(\bm{G}),
\end{equation}
 where
 $  \mathsf{V}_\text{conc}(\bm{G}) = \sigma_\text{conc,rel}^2\,\bm{G}\,\bm{G}^T$  has the same tensor product structure as $\mathsf{V}_\text{conc}(\bm{\alpha})$ in \eqref{eq:variance_model_concentration}.
In contrast, $\mathsf{V}_\text{noise}(\bm{G}) = \mathsf{K} \, \mathsf{V}_\text{noise}(\bm{\alpha}) \, \mathsf{K} ^T$ is,  unlike $\mathsf{V}_\text{noise}(\bm{\alpha})$, not 
diagonal,  because of the non-locality of the KK transform.

\subsection{Uncertainty Propagation in Linear Fit}
The covariance matrix $ \mathsf{V}(\bm{y})$ of 
\begin{equation}
 \bm{y} = \bm{B}^* - \bm{B}_0 = \bm{B}^* - \bm{G}_\text{VIS}  - \bm{G}_\text{UV} - \bm{G}_{\rm H_2O}
\end{equation}
is obtained by
\begin{align}
 \begin{split}
 \mathsf{V}(\bm{y}) &= \mathsf{V}(\bm{B}^*) + \mathsf{V}(\bm{B}_0) \\
 &= \mathsf{V}(\bm{B}^*) + \mathsf{V}(\bm{G}_\text{VIS}) +  \mathsf{V}(\bm{G}_\text{UV})  + \mathsf{V}(\bm{G}_{\rm H_2O}). \label{eq:Vy}
 \end{split}
\end{align} 
This decomposes into a noise and a concentration term $\mathsf{V}(\bm{y}) = \mathsf{V}(\bm{y})_\text{noise}+\mathsf{V}(\bm{y})_\text{conc}$.
For the contributions from concentration uncertainties, one finds (cf. \eqref{eq:Vy} and \eqref{eq:variance_concentration_error})
\begin{align}
 \mathsf{V}_\text{conc}(\bm{y}) 
 = \sum_{j=1}^4  \bm{v}_j\,  \bm{v}_j^T.
 \label{eq:Vyconc}
\end{align}
with
\begin{align}
\begin{aligned}
 \bm{v}_1:=%\bm{u}_\text{conc}(\bm{B}^*)  	  =  
    \sigma_\text{conc,rel} \, \bm{B}^*, {}&{\quad} 
 \bm{v}_2:= %\bm{u}_\text{conc}(\bm{G}_\text{VIS})=
    \sigma_\text{conc,rel} \, \bm{G}_\text{VIS}, \\
 \bm{v}_3:=%\bm{u}_\text{conc}(\bm{G}_\text{UV})=
    \sigma_\text{conc,rel} \, \bm{G}_\text{UV}, {}&{\quad}
 \bm{v}_4:=%\bm{u}_\text{conc}(\bm{G}_{\rm H_2O}) =
    \frac{u_{\rho_{\rm Hb}}}{\rho_{\rm Hb}} \, \bm{G}_{\rm H_2O}.
\end{aligned}
\end{align}
Similarly, we obtain $\mathsf{V}_\text{conc}(\bm{B}^*) = \bm{v}_1\, \bm{v}_1^T$.

However, for weighting the linear fit, we  do not use this covariance matrix, but only the noise terms, \ie, $\mathsf{V} = \mathsf{V}_\text{noise}(\bm{y})$ with
\begin{align}
 \mathsf{V}_\text{noise}(\bm{y})
 &= 
 \mathsf{V}_\text{noise}(\bm{B}^*) + \mathsf{K}\,\mathsf{V}_\text{noise}(\bm{\alpha}_\text{VIS})\,\mathsf{K}^T + \mathsf{K}\,\mathsf{V}_\text{noise}(\bm{\alpha}_\text{UV})\,\mathsf{K}^T
\end{align}
(noise in $n_{\rm H_2O}$ is negligible).
{Here $\mathsf{V}_\text{noise}(\bm{B}^*)$ is a diagonal matrix containing the uncertainties given for the refractive increment in \cite{Friebel2006modelfunction}.}
The reason to use $\mathsf{V} = \mathsf{V}_\text{noise}(\bm{y})$ is that the tensor-product structure (cf. \eqref{eq:Vyconc} and \eqref{eq:variance_concentration_error}) of the
systematic covariance matrices makes them singular and hence the full covariance matrix (including noise) close to singular, which is a problem for the numerics.
But after all, weighting with uncertainties that arise from a global prefactor for all values has
little sense.

Since $\hat{\bm{f}} = \mathsf{F}\,\bm{y}$ (\eqref{eq:fit_fFy}) is obtained by linear transformation, the covariance matrix is formally obtained as
\begin{equation}
 \mathsf{V}_\text{noise}(\hat{\bm{f}}) =  \mathsf{F}\, \mathsf{V}\,\mathsf{F}^T = \mathsf{H}\,(\mathsf{H}^T\mathsf{V}^{-1}\mathsf{H})^{-1}\mathsf{H}^T.%,
\end{equation}
% where the form in the last step is to be preferred to obtain numerical values, but solving systems of equations instead of using the inverse matrices explicitly.
For the concentration contributions one obtains
\begin{align}
 \mathsf{V}_\text{conc}(\hat{\bm{f}}) &= \sum_{j=1}^4 \bm{w}_j\, \bm{w}_j^T \text{\quad{}with\quad}\bm{w}_j = \mathsf{F}\, \bm{v}_j.
\end{align}

Although $\hat{\bm{f}}$ is formally the result of the fit, the quantity we are interested in is not  $\hat{\bm{f}}$ but rather 
\begin{equation}
 \hat{\bm{B}} = \bm{B}_0 + \hat{\bm{f}} = \bm{B}_0 + \mathsf{F}\,\bm{y}  =  \bm{B}_0 + \mathsf{F}\,(\bm{B}^* - \bm{B}_0) = (\mathsf{F} - \mathbb{1})\, \bm{B}_0  + \mathsf{F}\,\bm{B}^*,
\end{equation}
where $\mathbb{1}$ denotes the identity matrix.
We can assume $\bm{B}^*$ (the measured refractive increment) and $\bm{B}_0$ (the part of the refractive increment computed from absorption spectra and water RI) to be
uncorrelated  and obtain
%, at least concerning detector noise,
we obtain
\begin{align}
\begin{split}
 \mathsf{V}(\hat{\bm{B}}) 
    &= (\mathsf{F} - \mathbb{1})\, \mathsf{V}(\bm{B}_0)\, (\mathsf{F}^T - \mathbb{1}) + \mathsf{F} \,\mathsf{V}(\bm{B}^*)\,\mathsf{F}^T \\
    &= \mathsf{V}(\bm{B}_0) + \mathsf{V}(\hat{\bm{f}}) - \mathsf{F}\,\mathsf{V}(\bm{B}_0) - \mathsf{V}(\bm{B}_0)\,\mathsf{F}^T. \label{eq:VB}
\end{split}
\end{align}
This also decomposes into
\begin{equation}
 \mathsf{V}(\hat{\bm{B}}) = \mathsf{V}_\text{noise}(\hat{\bm{B}}) +  \mathsf{V}_\text{conc}(\hat{\bm{B}}),
\end{equation}
where both parts are computed separately. $\mathsf{V}_\text{conc}(\hat{\bm{B}})$ contributes stronger to the diagonal elements of $\mathsf{V}(\hat{\bm{B}})$ (\ie, the variances of the $\hat{B}_i$) than 
$\mathsf{V}_\text{noise}(\hat{\bm{B}})$. On the other hand,$\mathsf{V}_\text{conc}(\hat{\bm{B}})$, is  strongly correlated among all elements, whereas
$\mathsf{V}_\text{noise}(\hat{\bm{B}})$ is not. To illustrate the degree of correlation, the numerical derivative $B'(\lambda_i)$, obtained from finite 
differences, is plotted in Fig.~\ref{fig:B_prime} with the corresponding standard deviations.

\subsubsection{Deoxyhemoglobin}
The refractive increment for deoxyhemoglobin in \eqref{eq:Bdeoxy} can  also be written as
\begin{align}
 \bm{B}^\text{deoxy}% &= \bm{B}_0^\text{deoxy} + \hat{\bm{f}} = \bm{B}_0^\text{deoxy} + \mathsf{F}\,(\bm{B}^*- \bm{B}_0)\nonumber \\
 &= 
  \bm{G}_\text{lit}^\text{deoxy} - \mathsf{F}\,\bm{G}_\text{lit} - (\mathsf{F}- \mathbb{1})\,\bm{G}_{\rm H_2O} + \mathsf{F}\,\bm{B}^*,
\end{align}
from which follows
\begin{align}
\begin{split}
 \mathsf{V}_\text{noise}(\bm{B}^\text{deoxy}) ={}&  \mathsf{V}_\text{noise}(\bm{G}_\text{lit}^\text{deoxy})\\
 &{}+
  \mathsf{F}\, \mathsf{V}_\text{noise}(\bm{G}_\text{lit}) \, \mathsf{F}^T + 
  \mathsf{F}\, \mathsf{V}_\text{noise}(\bm{B}^*) \, \mathsf{F}^T
  \end{split}\\
  \mathsf{V}_\text{conc}(\bm{B}^\text{deoxy}) ={}& \sum_{j=1}^6 \bm{u}_j\, \bm{u}_j^T
\end{align}
with
\begin{align}
 \begin{aligned}
  \bm{u}_j := \mathsf{F}\,\bm{v}_j \text{ for } j=1,2,3; {}&{\quad} \bm{u}_4 := (\mathsf{F}- \mathbb{1})\,\bm{v}_4
  \\
  \bm{u}_5 :=  \sigma_\text{conc,rel} \, \bm{G}^\text{deoxy}_\text{VIS}, {}&{\quad}
  \bm{u}_6 :=     \sigma_\text{conc,rel} \, \bm{G}^\text{deoxy}_\text{UV}.
 \end{aligned}
\end{align}

\subsection{Peptide Absorption: Influence of Peak Shape}\label{sec:peak_shape}
We have assumed the exact shape of the peptide peak to be of minor importance, such that the results depend mainly on its position, total strength and width.
In order to justify this assumption quantitatively, we evaluated an alternative model, where the peak has a rectangular shape.
% The analytic expression for the KK transformation is given in appendix~\ref{app:trafo_Lorentzian}. %% is not given there anymore

The comparison between results with the Lorentzian peak centered at $L=\SI{187}{nm}$ and half width at half maximum $\Gamma = \SI{11.6}{nm}$
and a rectangle centered around $L_\Pi = L$ and half width $\Gamma_\Pi = \Gamma$  reveals that deviations between the two models are
negligible in comparison to the propagated data uncertainties for the majority of wavelengths. Only at wavelengths $\lambda < \SI{400}{nm}$
does the deviation $|\hat{B}(\lambda)- \hat{B}_\Pi(\lambda)|$
{ exceed the estimated uncertainties due to noise, but is still smaller than the total uncertainty including concentration errors.
For $\lambda>\SI{400}{nm}$ the deviation is at least one order of magnitude smaller than the total uncertainties.}

\section*{Acknowledgment}
The authors would like to thank Moritz Friebel for providing in tabulated form the absorbance data for  oxygenated and deoxygenated  hemoglobin published in \cite{Friebel2005concentratedHb} as figures.

% Bibliography
\bibliography{Hb_RI_KK_sub2_arXiv}

% Full bibliography added automatically for Optics Letters submissions
% Note that this extra page will not count against page length
\ifthenelse{\equal{\journalref}{ol}}{%
\clearpage
\bibliographyfullrefs{sample}
}{}

\end{document}